# From Many, One: Genetic Control of Prolificacy during Maize Domestication

David M. Wills[a], Clinton Whipple[b], Shohei Takuno[c], Lisa E. Kursel[a], Laura M. Shannon[a], Jeffrey Ross-Ibarra[c,d], and John F. Doebley[a]

[a]Department of Genetics, University of Wisconsin-Madison, Madison, WI 53706; [b]Department of Biology, Brigham Young University, Provo, UT 84602 ; [c]Department of Plant Sciences, University of California, Davis CA 95616; [d]The Genome Center, and Center for Population Biology, University of California, Davis CA 95616.

**Corresponding Author:** John Doebley, Department of Genetics, University of Wisconsin-Madison, Madison, WI 53706, voice: 608-265-5803; email: jdoebley@wisc.edu

**Classification:** Biological Sciences – Agricultural Sciences, Plant Biology

**Abstract**

A reduction in number and an increase in size of inflorescences is a common aspect of plant domestication.  When maize was domesticated from teosinte, the number and arrangement of ears changed dramatically.  Teosinte has long lateral branches that bear multiple small ears at their nodes and tassels at their tips.  Maize has much shorter lateral branches that are tipped by a single large ear with no additional ears at the branch nodes.  To investigate the genetic basis of this difference in prolificacy (the number of ears on a plant), we performed a genome-wide QTL scan.  A large effect QTL for prolificacy (*prol1.1*) was detected on the short arm of chromosome one in a location that has previously been shown to influence multiple domestication traits.  We fine-mapped *prol1.1* to a 2.7 kb interval or "causative region" upstream of the *grassy tillers1* gene, which encodes a homeodomain leucine zipper transcription factor.  Tissue *in situ* hybridizations reveal that the maize allele of *prol1.1* is associated with up-regulation of *gt1* expression in the nodal plexus.  Given that maize does not initiate secondary ear buds, the expression of *gt1* in the nodal plexus in maize may suppress their initiation.  Population genetic analyses indicate positive selection on the maize allele of *prol1.1*, causing a partial sweep that fixed the maize allele throughout most of domesticated maize.  This work shows how a subtle *cis*-regulatory change in tissue specific gene expression altered plant architecture in a way that improved the harvestability of maize.

**Significance Statement:**  Crop species underwent profound transformations in morphology during domestication.  Among crops, maize experienced a more striking change in morphology than other crops. Among the changes in maize from its ancestor, teosinte, was a reduction in the number of ears per plant from 100 or more in teosinte to just one or two in maize.  We show that this change in ear number has a relatively simple genetic architecture involving a gene of large effect, called *gt1*.  Moreover, we show that *gt1* experienced a tissue-specific gain in expression, demonstrating how simple changes in genes can lead profound differences in form.

**Introduction**

The ''domestication syndrome'' of crop plants is a suite of adaptive traits that arose in response to direct and indirect selection pressures during the domestication process (1-3). This suite of traits includes an increase seed or fruit size, larger inflorescences, an increase in apical dominance, more determinate growth and flowering, loss of natural seed dispersal, loss of seed dormancy, and gain of self-compatibility. These traits make crop plants easier to cultivate and harvest, resulting in increased value for human use.

Among the domestication syndrome traits, the increase in apical dominance improves agricultural performance by enhancing harvestability. Apical dominance confers a reduction in the number of branches and inflorescences per plant. The inflorescences that do form, however, have either more and/or larger fruits or seeds. Thus, increased apical dominance can afford easier harvestability by reducing the number of inflorescences to be harvested without a concomitant loss in yield per plant. Moreover, larger seeds allow for more vigorous growth after germination when seedlings can face intense competition from weedy species. Finally, the fewer but larger inflorescences mature in a narrower window of time, enabling all the fruit/seed of a plant to be harvested at the same time of optimal maturation.

Maize was domesticated from Balsas teosinte (*Zea mays* subsp. *parviglumis*) through a single domestication event in Mexico about 9000 years ago (4, 5). During maize domestication, there was a profound increase in apical dominance such that the amount of branching and the number, size and arrangement of the female inflorescences (ears) changed dramatically (6, 7). The teosinte plant has multiple long lateral branches, each tipped with a tassel. At each node along these lateral branches, there are clusters of several small ears (Fig. 1A). Summed over all branches, a single teosinte plant can easily have more than 100 small ears. By comparison, the maize plant has relatively few lateral branches (often just two), each tipped by a single large ear rather than a tassel as in teosinte (Fig. 1C). Modern commercial varieties of maize typically have only one or two ears per plant, and even traditional landraces of maize rarely have more than 6 ears per plant. In maize genetics and breeding, the number of ears on a plant is scored as *prolificacy*, teosinte having high and modern maize low prolificacy.

Here, we report a genome-wide scan for prolificacy QTL using a maize-teosinte $BC_2S_3$ mapping population (8). We detected eight QTL, including one of large effect on the short arm of chromosome 1. We fine-mapped this QTL to a 2.7 kb "causative region" located 7.5 kb upstream of the coding sequence of the known maize gene *grassy tillers1* (*gt1*), which encodes a homeodomain leucine zipper (HD-ZIP) transcription factor (9). Expression assays indicate that the causative region has little effect on the overall abundance of the maize vs. teosinte transcripts, however, tissue *in situ* hybridizations show that the maize allele of *prol1.1* is associated with a tissue specific change *gt1* expression. Molecular population genetic analysis suggests that the causative region was the target of a partial selective sweep that brought a haplotype at low frequency in teosinte to a higher frequency over most of the range of maize landraces. Within the causative region, the common maize haplotype is distinguished from the common teosinte haplotype by multiple polymorphisms including several transposable element insertions. Our results how a subtle change in the tissue specific gene expression altered maize plant architecture in a way that improved the harvestability.

## Results

**A major QTL (*prol1.1*) largely controls prolificacy.** Whole genome QTL mapping for loci affecting prolificacy was performed using a set of 866 maize-teosinte $BC_2S_3$ recombinant inbred lines (RILs). This analysis identified eight QTL, distributed across the first 5 chromosomes (Fig. 2, Table 1). Of the eight QTL, one has a much larger effect than the other seven. This QTL (*prol1.1*) is located on the short arm of chromosome 1 and accounts for 36.7% of the phenotypic variance. Plants in the mapping population that are homozygous teosinte at *prol1.1* typically produce multiple ears at each node like teosinte (Fig. 1B). The 1.5 LOD support interval surrounding *prol1.1* defines a 0.79 Mb segment between 22.63 Mb and 23.42 Mb (B73 Reference Genome v2) on chromosome 1. This region contains just 25 annotated genes including *gt1*. The other seven QTL have much smaller LOD scores and much smaller effects. This disparity in QTL size suggests that although the seven smaller QTL contribute to prolificacy, the phenotype is primarily controlled by *prol1.1*.

***prol1.1* maps to the promoter of *gt1*.** We chose *prol1.1* for fine-mapping to identify the underlying causative gene. Two markers (*umc2226* and *bnlg1803*) that flank the QTL interval were used to screen for recombinant chromosomes in one of the 866 $BC_2S_3$ RILs that is heterozygous in the *prol1.1* QTL interval. After screening ~4000 plants of this RIL, 23 plants with a cross-over between the two markers were identified and self-pollinated to create progeny lines homozygous for the 23 recombinant chromosomes. The physical position of each of the 23 recombination events was determined using a combination of gel-based markers and DNA sequencing (Figs. 3, S1; Table S1).

Progeny lines homozygous for the 23 recombinant chromosomes were grown in a randomized-block design and scored for prolificacy. We also included two lines derived from the same $BC_2S_3$ RIL as controls: one homozygous teosinte and the other homozygous maize in the QTL interval. This set of 25 progeny lines fell into two discrete classes for prolificacy (Fig. 3). One class, which included the maize control line, had an average prolificacy score of $2.38 \pm 0.05$ ears. The other class, which included the teosinte control line, had an average prolificacy score of $7.24 \pm 0.12$ ears. Separately, to estimate dominance relationships, we compared the trait values of the maize, teosinte and heterozygous genotypic classes at *prol1.1* The dominance/additivity ratio is 0.08, indicating additive gene action (Table S2).

Examination of the relationship between the two phenotypic classes and the recombination breakpoints revealed that all members of the maize class carry maize chromosome between markers SBM07 (AGP v2: 23,232,048) and SBM08 (AGP v2: 23,234,775) (Fig. 3, S1). Correspondingly, all members of the teosinte phenotypic class carry teosinte chromosome between these two markers. No other chromosomal region shows this absolute correspondence with phenotype. Thus, substitution mapping based on the recombination breakpoints indicates that *prol1.1* or the factor that governs prolificacy maps to this interval. This interval, which we will refer to as the "causative region," is approximately 7.5 kb upstream of *gt1* and measures 2720 bp in W22, 3142 bp in our teosinte parent, and 2736 bp in the B73 reference genome (Fig. 3, S1). The sequence alignment of W22 and the teosinte parent expands to ~4.2 kb because there are several large insertions unique to either W22 or teosinte (see below).

**The decrease in prolificacy in maize is correlated with an increase in kernel weight.** The maize allele of *prol1.1* confers a reduction in ear number, which by itself would cause a reduction in yield. To test whether there is a compensatory increase in either the number of kernels per ear or kernel weight, we assayed plants of the $BC_2S_3$ family used for fine-mapping to determine if *prol1.1* has associated effects on these traits. The *prol1.1* maize allele is not associated with an increase in ear size as measured by the total number of spikelets (kernel forming units) produced in the primary ear (maize = 418, heterozygous = 423, teosinte = 421,

P=0.86; Table S2).  However, the maize allele is associated with an increase in kernel weight (maize = 0.216 g, heterozygous = 0.208 g, teosinte = 0.187 g, P<0.0001; Table S2).  Other aspects of plant architecture such as tillering and the number of nodes along the maize culm that produce ears do not appear to be affected by *pro1l.1* (Table S2).  Thus, these data suggest that the reduction in secondary ears caused by *pro1l.1* in maize was compensated for by an increase in kernel weight such that yield itself may not have changed.  To confirm this interpretation would require a formal yield trial comparing the maize and teosinte genotypes.

**Maize and teosinte alleles of *gt1* show near equal expression.**  The location of *pro1l.1* at ~7.5 kb upstream of coding sequence of *gt1* suggests that it may represent a *cis*-regulatory element of *gt1*.  To investigate this possibility, we used ESTs from Genbank and genomic sequence of our maize and teosinte parents to construct a gene model for *gt1* (Fig. S2).  This model agrees with the *gt1* gene model presented elsewhere (9).  *gt1* possesses three exons with two small introns and a transcript of ~1350 bp that encodes a protein of 239 amino acids.  The homeodomain and a putative nuclear localization signal are located in Exon 2.

We performed RT-PCR with primers designed to amplify most of the predicted transcript (1203 bp of the predicted 1350 bp) using cDNAs isolated from immature ear-forming axillary branches of isogenic lines derived from our mapping population possessing the maize and teosinte alleles.  We observed three size classes of RT-PCR products, presumably corresponding to three splice variants or isoforms of *gt1* (Fig. 4).  The three size classes are present with both maize and teosinte alleles.  We cloned and sequenced all three size classes and aligned these with the genomic sequence (Fig. S3).  The largest class contains the entire predicted open reading frame, encoding a predicted protein of 239 amino acids.  The middle-sized product is missing most of Exon 2 and part of Exon 3.  The smallest-sized product is missing all of Exon 2 and parts of Exons 1 and 3.  Critically, the middle and small-sized products are both missing the homeodomain and all or part of the putative nuclear localization signal.

The relative band intensities of different sized RT-PCR products (Fig. 4) suggest that transcript abundance for the isoforms differs between the maize and teosinte alleles: teosinte having a greater abundance of the full length product and maize a greater abundance of the middle-sized product that lacks the homeodomain. To test whether these differences in band intensity for the different isoforms are independent of the causative region, we performed RT-PCR with two of our recombinant isogenic lines. One of these has the teosinte causative region linked to the maize coding sequence (T:M), and the other has the maize causative region linked to the teosinte coding sequence (M:T). RT-PCR assays with these recombinant lines confirm that the differential band intensity for the isoforms is determined by the coding sequence and not the causative region 7.5 kb upstream of the coding sequence (Fig. 4).

To investigate the effect of the causative region on transcript abundance for our maize and teosinte alleles, we used an allele specific expression assay (10). cDNA was made from RNA from immature ear-forming axillary branches of plants heterozygous at *prol1.1-gt1*. PCR primers were designed flanking a 2 bp indel in the 3' non-translated region that distinguishes the maize and teosinte alleles (Fig. S2). This indel is in all three isoforms, and thus PCR products measure the overall difference in the abundance of the maize and teosinte transcripts without regard to any differences in relative abundance of the isoforms between maize and teosinte. In a heterozygous plant, the maize and teosinte alleles are expressed in the same cells with a common set of trans-acting factors, therefore any difference in transcript abundance of the alleles in heterozygous plants must be due to *cis*-regulatory factors. This assay shows a ratio of 1.35 teosinte:maize *gt1* transcript, suggesting a modest but statistically significant excess of teosinte relative to maize transcript (z-test, P<0.001).

As an additional test of the effects of the causative region on *gt1* transcript abundance, we used quantitative PCR (qPCR) to compare overall *gt1* transcript abundance in immature ear-forming axillary branches of isogenic lines that are homozygous for the maize vs. teosinte alleles at *prol1.1-gt1*. For this assay, we used a primer pair in the 3' UTR of all three isoforms. The abundance of *gt1* transcript relative to *actin* transcript for the teosinte class (1.03, *n*=12) was slightly higher than the maize class (0.88, *n*=12), however this difference is not statistically significant (t-test, P=0.077). Both the allele specific expression assay and qPCR suggest that the teosinte transcript abundance might be slightly higher than that of maize, but any difference is modest.

**Maize *prol1.1* directs increased *gt1* expression in primary branch nodes.** Although a substantial change in *gt1* transcript levels was not detected between the maize and teosinte alleles of *prol1.1* in immature ear-forming axillary branches, we hypothesized that the absence of secondary ears in maize could be caused by a more subtle change that does not drastically alter overall transcript level but instead impacts the domain of *gt1* expression. In order to test for such a tissue-specific expression difference, we performed RNA *in situ* hybridization on immature primary ear-forming branches of lines containing all possible combinations of the maize and teosinte causative region (*pro1.1*) and *gt1* coding sequence (M:M, T:T, M:T, and T:M).  A previous study demonstrated that *gt1* is strongly expressed in the leaves of dormant tiller-forming lateral buds (9), thus we anticipated that *gt1* expression might differ in the leaves (husks) surrounding secondary ear buds of maize and teosinte.  Contrary to this expectation, our sections revealed that lines containing the maize allele of *pro1.1* (M:M and M:T) rarely, if at all, initiate secondary ear buds (*SI Appendix*).  Expression of *gt1* was observed in young leaves surrounding secondary ears of lines containing the teosinte allele of *pro1.1* (T:T and T:M) (Fig. S4), but was weak compared to dormant buds (9), and required an extended incubation for detection, suggesting that these secondary ears are not dormant.  Interestingly, an up-regulation of *gt1* expression was observed in the stem node or nodal plexus (11) of primary branches for lines containing the maize allele of *pro1.1* (M:M and M:T, Fig. 5 A,B).  This nodal *gt1* expression was either absent or only weakly detectable above background in lines containing the teosinte allele of *pro1.1* (Fig. 5 C,D).  While the nodal stripe of *gt1* was weak, the difference between the maize and teosinte *pro1.1* lines was consistently observed.  Taken together, these observations suggest that the allelic differences at *pro1.1* involve changes in a *cis*-regulatory element that causes increased *gt1* expression in the nodal plexus of maize, which in turn inhibits the initiation of secondary ear buds.

**A partial selective sweep occurred at *pro1.1*.**  To investigate whether the causative region shows evidence of past selection during maize domestication, we sequenced the entire causative region (~2.7 kb) plus flanking sequence (~1000 bp upstream and ~700 bp downstream) in 15 inbred maize landraces and 9 inbred teosinte (Table S4).  Diversity statistics across the region in both teosinte ($S = 85$, $\pi = 0.00844$ and Tajima's $D = -1.16$) and maize ($S = 32$, $\pi = 0.00307$ and Tajima's $D = -0.439$) are within the previously estimated range of these statistics for neutral genes (12), where $S$ and $\pi$ were the number of segregating sites and nucleotide diversity, respectively.  Although these data would superficially appear to be consistent with a loss of diversity due to the domestication bottleneck alone, a neighbor-joining tree of the sequences separates most maize from most teosinte sequences in the causative region (Fig. S5A).  This separation of the mostly maize and mostly teosinte clusters reflects differences at numerous SNPs and multiple putative transposon insertions (Fig. S6).  We will refer to these maize and teosinte clusters hereafter as the class-*M* and class-*T* haplotypes, respectively.  Linkage disequilibrium (LD) analysis of maize sequences confirms this separation, identifying a 2.5 kb block of strong LD corresponding to SNPs that differentiate class-*M* from class-*T* maize sequences (Fig. 6A, S7).  This high LD block lies completely within the 2.7 kb causative region. The maize class-*M* haplotype in this block exhibits extremely low levels of nucleotide diversity ($\pi = 0.000740$) and a strongly negative Tajima's *D* value ($D = -1.966$).  These values are extremely unlikely under neutrality (p<0.01; *SI Appendix*), leading us to investigate instead a partial sweep model to explain the observed sequence data.

To investigate the unusual pattern of diversity for the maize class-*M* haplotypes, we applied a maximum likelihood method to estimate the selection coefficient (*s*) and the degree of dominance (*h*) using structured coalescent simulations (*SI Appendix*).  We specified a partial sweep model (Fig. 6B), consistent with the observation of both class-*M* and class-*T* haplotypes in domesticated maize sequences, and performed structured coalescent simulations over a wide range of parameter settings similar to previous studies (12, 13).  Our maximum likelihood estimates suggest that the class-*M* allele is dominant ($h = 1.0$) and under reasonably strong selection ($s = 0.0015$) (Fig. 6C).  We also estimated the age of class-*M* haplotype to be ~13,000 generation ago using Thomson's method (14, 15). Although the observed length (2.5 kb) of the swept region may seem short, simple calculations show that this length falls within the ~1-7 kb range expected given available estimates of recombination and the age of the haplotype (*SI Appendix*).

We assayed a diverse sample of maize and teosinte to better estimate the frequencies of the class-*M* and class-*T* haplotypes.  We used an ~250 bp insertion specific to the class-*T* haplotype as a marker.  We observed that the class-*M* haplotype exists at a relatively low frequency in ssp. *parviglumis* (5%) and ssp. *mexicana* (8%) while the class-*T* haplotype exists at a moderate frequency in maize landraces (29%) (Table 2).  These frequencies are consistent with the partial selective sweep discussed above that brought the class-*M* haplotype from a low frequency (5%) in the progenitor population to a relatively high frequency (71%) in domesticated maize.

An examination of the distribution of the class-*T* haplotype in maize shows a distinct geographic pattern (Fig. S8).  With only three exceptions, the class-*T* haplotype is limited to southern Mexico, the Caribbean Islands and the northern coast of South America.  One exception is its occurrence in the landrace Tuxpeño Norteño in northern Mexico, but this is a landrace thought to be recently derived from the landrace Tuxpeño of southern Mexico (4). The two other exceptions are found in southern Brazil in landraces thought to have been brought to Brazil in the 1800s from the southern USA (16). In turn, the southern US landraces are thought to have been brought there from southern Mexico and the Caribbean in the 1600s by the Spanish (17). Thus, the class-*T* haplotype in maize has a distribution centered on southern Mexico and the Caribbean with recent dispersals to other regions.

**Discussion**

A critical challenge during the domestication of crop plants was to improve the harvestability of the crop as compared to its progenitor. Wild species are adapted to "spread their bets" and thereby increase the probability of successful reproduction under diverse environments (2). In unfavorable environments, wild plants can flower and mature rapidly, producing smaller numbers of branches, inflorescences, flowers and seeds but still complete their reproductive cycle. In favorable environments, wild plants can flower over a longer period, sequentially producing more branches, inflorescences, flowers and seeds over time, maximizing their reproductive output. The latter strategy is not optimal for a crop as greater efficiency of harvest is achieved by having all seed mature synchronously. Similarly, harvesting a single large inflorescence or fruit from a plant is easier than harvesting dozens of smaller ones (18). Thus, diverse crops have been selected to produce smaller numbers of larger seeds, fruits or inflorescences as a means of improving harvestability (2). In the terminology of modern day maize breeders, crops were selected to be less prolific.

Our QTL mapping for prolificacy confirms the results of three prior studies that indicated this trait is controlled by a relative small number of QTL including one of large effect on the short arm of chromosome 1. First, in an $F_2$ cross of Chalco teosinte (*Zea mays* ssp. *mexicana*) with a Mexican maize landrace (Chapalote), four prolificacy QTL were detected (19). One of these QTL, located on the short arm of chromosome 1, had a large effect, accounting for 19% of the phenotypic variance. Second, in an $F_2$ cross of Balsas teosinte with a different Mexican maize landrace (Reventador), seven prolificacy QTL were detected (20). One of these QTL, again located on the short arm of chromosome 1, had a large effect, accounting for 25% of the phenotypic variance. Finally, in a maize-teosinte $BC_1$ cross of Balsas teosinte by a US inbred line (W22), seven prolificacy QTL were detected (21).  All seven QTL had small effects, but the one that explained the greatest portion of the variance (4.5% averaged over two environments) was on the short arm of chromosome 1.  As in these prior studies, the QTL mapping reported here indicates that prolificacy is under relatively simple genetic control, involving only 8 QTL but including one QTL (*prol1.1*) of large effect.  *prol1.1* accounted for 36.7% of the variation in the number of ears and reduces the number of ears from 7.2 for teosinte homozygous class to 2.4 for the maize homozygous class.

The genetic architecture of the change in prolificacy during domestication appears to be relatively simple in several other crops as well.  In tomato, five QTL of roughly equal effects for the number of flowers per truss between wild and domesticated tomato were detected (22, 23).  In the common bean*,* three QTL were detected for the reduction in the number of pods per plant in a cross of wild and domesticated bean (24).  The QTL of largest effect confers a reduction from 29 to 17 pods per plant and accounts for 32% of trait variation. In pearl millet, the reduction in the number of spikes per plant is governed by four QTL, including one that controls 37% of trait variation (25).  In sunflower, the reduction of number of heads per plant was governed by seven QTL, one of which had a much large effect than the other six (26).  This large effect QTL accounts for a difference of 4.8 heads per plant between the cultivated and wild genotypes, and it co-localizes with the previously described *Branching* (*B*) locus, which is known to influence apical dominance (27). Thus, simple genetic architecture including QTL of relatively large effect is common for this trait.

One theory of crop domestication is that domestication traits are often the result of recessive, loss of function alleles (28).  Contrary to this expectation, *prol1.1* acts in an additive fashion with a dominance/additivity ratio of 0.08, suggesting that domestication did not involve selection for a simple loss of function. Moreover, our expression assays indicate that *gt1* has roughly equal expression in maize and teosinte ear-forming axillary branches and the phenotypic change is caused by a relatively subtle gain/increase of expression in the nodal plexus of the ear-forming branches of maize. These results demonstrate that rather than a simple loss of function allele, the gene underlying this QTL experienced an increase or gain of expression in a specific tissue. While selection for loss of function alleles may be a common feature of domestication, none of the three positionally mapped maize domestication QTL (*teosinte branched1*, *teosinte glume architechture1*, and *gt1*) involved a loss of function allele (29, 30, this paper).

Seventy-five years ago, the "teosinte hypothesis" that a small number of large effect genes substitutions could convert teosinte into a useful food crop was proposed (31). The experimental basis for this model was that maize-like and teosinte-like segregants were recovered in a large maize-teosinte $F_2$ population at frequencies, suggesting that as few as five loci might control the critical differences in ear architecture. Subsequent QTL mapping identified six regions of the genome that harbor QTL of large effect on plant and ear architecture, consistent with the teosinte hypothesis (32). Fine-mapping of two of these QTL identified an underlying gene of large effect in both cases. One of these is *teosinte glume architecture* (*tga1*) that controls the difference between covered vs. naked grain (30), and the other is *teosinte branched* (*tb1*), which conferred increased apical dominance during domestication (29). In this paper, we have shown that a gene of large effect (*gt1*) also underlies a third of these six QTL of large effect. This result adds further support to the view that a small number of genes of large effect were key in the dramatic morphological changes that occurred during maize domestication. Nevertheless, it is also clear a larger number of QTL of smaller effect on morphology were also involved in converting teosinte into modern maize (8, 32, 33).

The role played by genes of large effect, like *gt1*, is not limited to maize domestication, but seems to be a common feature of plant domestication (34). Recently, a large effect gene in sorghum that encodes a YABBY transcription factor was shown to control shattering vs. non-shattering inflorescences (35). Previously, two domestication genes controlling shattering had been identified in rice, one encoding a homeodomain and the other a myb-domain transcription factor (36, 37). In tomato, two domestication genes for increase in fruit size have been isolated, one encoding a YABBY transcription factor and the other a putative cell signaling gene (38, 39). A single gene (*PROG1*), which encodes a zinc finger transcription factor, controls differences in plant architecture and grain yield between wild and cultivated rice (40, 41).

The fine-mapping of *prol1.1* was initiated using a publically available set of maize-teosinte RILs (8). These RILs allow some QTL to be mapped to relatively small intervals. We mapped *prol1.1* to a 0.79 Mbp segment that included only 25 annotated genes and then fine-mapped it to a 2.7 kbp causative interval. These same maize-teosinte RILs were recently used fine-map a QTL (*dtp10.1*) for photoperiod response that was involved in the adaptation of maize to northern latitudes (8, 42). The *dtp10.1* QTL was mapped to a 7.6 Mbp interval containing 103 annotated genes, and then fine-mapped to a 202 kbp interval containing a single annotated gene (*ZmCCT*). Features of *prol1.1* and *dtp10.1* that made them good candidates for fine-mapping were (1) having large effects with strong statistical support (LOD>100) so that progeny lines with recombinant chromosomes possessing the maize vs. teosinte alleles of the QTL segregated into two distinct classes (i.e. Mendelized) and (2) being located in genomic regions with sufficient recombination to capture multiple cross-overs per gene in an $F_2$ family of 2000 plants. For example, *prol1.1* is located near the end of the short arm of chromosome 1, where we observed a recombination rate $1.3 \times 10^{-3}$ cM/kbp which is over twice the genome-wide rate reported for a maize-teosinte crosses (21).

The location of *prol1.1* just 7.5 kb 5' of *grassy tillers1* (*gt1*) suggested that it may act as a *cis*-regulatory element of *gt1*. Whipple et al (9) identified *gt1* as a HD-Zip transcription factor, a class of proteins that is unique to plants. The role of *gt1* in maize development is complex. Although named for the excessive tillering caused by loss of function alleles, these alleles also cause the derepression of carpels in tassel florets, leading to the formation of sterile carpels (9). Additional changes include an increased numbers of ear-forming nodes along the main culm, elongation of the lateral branches, and elongation of the blades on the husk leaves. The formation of secondary ears is occasionally (but not typically) seen with maize *gt1* mutant allele consistent with the effect of *prol1.1* on *gt1* expression that we observed. The infrequency of this phenotype with the maize mutant alleles might be due to differences in genetic background between our lines for which about 10% of the genome comes from teosinte and the elite maize inbreds in which *gt1* mutant alleles have been assayed. One curiosity is that the teosinte allele we studied does not confer an increase in tillering (Table S2), suggesting the role of *gt1* in regulating tillering is conserved between maize and teosinte.

Another HD-Zip transcription factor, *six-rowed spike1* (*Vrs1*), has been identified as a domestication gene, controlling the change from two-rowed spikes in the wild progenitor of barley to six-rowed spikes found in domesticated barley (43). *Vrs1* is expressed in the lateral spikelet primordia of immature spikes of wild barley where it represses their development. Loss of function *vrs1* alleles selected during domestication fail to repress the development of these lateral spikelets, resulting in two additional fully fertile spikelets per rachis node. A comparison of *gt1* and *vrs1* offers an interesting contrast. Loss of function of *vrs1* alleles were selected in barley, producing a larger number of organs (spikelets or grains) per spike, while selection for an allele that confers the gain of nodal expression of *gt1* in maize caused a reduction in the number of organs (ears) per plant.  In maize, our data suggest the reduction in ear number may be compensated for by an increase in grain weight such that yield may not be affected.  It would be of interest to know if the production of more grains per spike in barley is compensated for by a reduction in the number of spikes per plant such that yield is not affected although harvestability is improved.

The nature of the causative polymorphism for *prol1.1* that governs *gt1* expression in the nodal plexus and represses secondary ear formation remains unknown. There are multiple polymorphisms that distinguish the class-*M* and class-*T* haplotypes for the causative region, all of which are potential candidates for the functional variant that controls expression in the nodal plexus (Fig. S6).  Among these polymorphisms are at least four transposable element insertions including Cinful, Pif/Harbinger, and hAT elements.  Given the evidence that a Hopscotch transposon is the functional variant at *tb1* (29), the transposons in the causative interval of *gt1* are good candidates for future functional assays.  Transposon inserts have also been identified in alleles of genes involved in millet and tomato domestication or improvement (44, 45), suggesting that transposons may be important contributors to regulator variation in crop plants.

DNA sequence analysis of the *prol1.1* locus in diverse maize and teosinte accessions revealed two distinct haplotypes.  Both haplotypes were present in maize and teosinte, but the class-*M* haplotype was common in maize and rare in teosinte.  Neutral coalescent simulations revealed that patterns of diversity in the class-*M* haplotype in maize were unlikely in the absence of selection, and subsequent parameter estimation supported a partial sweep model in which selection acted to increase the frequency of the class-*M* haplotype during domestication.  The estimated age of the class-*M* haplotype at 13,000 BP predates maize domestication and is consistent with its observed presence in about ~5% of the teosinte sampled.  This observation suggests that selection at *prol1.1* acted on standing variation, similar to observations for *tb1* (29) and *barren stalk1* (46).

It is curious that the class-*T* haplotype is found at a frequency of nearly 30% in maize, although the multi-eared phenotype that this haplotype confers is rare in maize.  Furthermore, none of the  maize races (Table S3) that carry the class-*T* haplotype are known to exhibit the multiple ears along a single shank.  These observations suggest that these landraces may have other factors that suppress the formation of multiple ears on a single shank.  Thus, there may have been two pathways to the switch from several to a single ear per node in maize, one governed by *prol1.1* and a second controlled by unknown factors that suppress multiple ear formation in plants carrying the class-*T* haplotype at *prol1.1*.

Previous analysis of *gt1* and surrounding sequence uncovered evidence of selection at the 3' UTR of the gene (9). We reanalyzed this sequence data (*SI Appendix*) and identified two distinct haplotypes distinguished by a ~40 bp indel. The class-*M* haplotype at this locus bears the signature of a partial sweep from standing variation similar to that seen at *prol1.1* (*SI Appendix*). A PCR survey of a large panel of maize landraces reveals that the class-*M* haplotype at the 3' UTR has an overall frequency of 78%. Combined with the small size of both sweeps and geographical differences in the abundance of each haplotype (Fig. S8), these results suggest that the class-*M* haplotypes at *prol1.1* and *gt1* may represent independent selective events (47), perhaps on different regulatory aspects of *gt1*. Neither *prol1.1* nor *gt1* were identified in a recent whole-genome analysis of selection during domestication (48), likely due to the short span of the selected region and the presence of the class-*T* allele in 30% of maize lines. This result highlights the difficulty in identifying small selected regions from genome-wide scans, especially in the case of soft sweeps (49,50).

The shade avoidance response in plants involves an increase plant height, a decrease in branching, reduction in the number of flowers, and early flowering (51). During domestication, human preference for easier harvestability resulted in a form of plant architecture that mimics the shade avoidance in that crops are less branched and produce fewer reproductive structures. Two maize domestication genes, *gt1* and *tb1*, are members of the developmental network controlling the shade avoidance response (9), suggesting that domestication acted to constitutively fix aspects of the shade avoidance syndrome in maize. As the shade avoidance network becomes better known, it will be of interest to see if additional genes within this network also play a role in domestication.

**Materials and Methods**

**QTL mapping.** Whole genome QTL mapping for prolificacy in maize was performed using a set of 866 maize-teosinte $BC_2S_3$ RILs that were genotyped at 19,838 markers using reduced representation sequencing (8, 52). In the cross, W22 was the recurrent parent and the teosinte parent was CIMMYT accession 8759 of *Zea mays* ssp. *parviglumis*. The 866 lines were grown in 2 blocks during summer 2009 and two additional blocks in summers 2010 and 2011 at the West Madison Agricultural Research Center in Madison, WI. All four blocks were randomized and contained 866 plots with 10 plants per plot. Prolificacy was treated as a binary trait; five plants per plot were scored as either having multiple visible ears on the primary lateral branch (1) or have a single visible ear on the primary lateral branch (0). Least Squared Means (LSMs) were determined for each line using the following model with PROC GLM (SAS Institute, Cary, NC):

Phenotype = Line + Seedlot(Line) + year + x-position(Block) + x-position * y-position (Block)

Line represents the RILs (1 through 866) and seedlot represents different seed productions for a single RIL. Year is 2009, 2010 or 2011, and for 2009 there were two blocks (A and B). The position of each plot within a block was recorded along the x-axis and y-axis of the field. Only the x-axis and the interaction between the x and y axes had a statistically significant effect so the y-axis was dropped from the model. The LSMs were used as the phenotypic values for QTL mapping. QTL mapping was carried out using multiple qtl mapping techniques in a modified version of R/qtl (53) that allows the program to take into account the $BC_2S_3$ pedigree of the lines (8).

**Fine mapping.** We used one of the $BC_2S_3$ RILs (MR0091) for fine-mapping of *prol1.1*. MR0091 is heterozygous for a 33.9 Mb region including this QTL and homozygous maize for all other prolificacy QTL. We screened ~4,000 MR0091-derived plants for cross-overs in the QTL interval between markers *umc2226* and *bnlg1803*. Twenty-three individuals with cross-overs in the QTL interval were identified and selfed. Selfed progeny from these 23 individuals that are homozygous for the recombinant chromosome plus two control lines (homozygous non-recombinant maize and teosinte) were grown in randomized block design with four blocks of 25 entries each. Prolificacy was scored as the total number of ears observed on the top two lateral branches of each plant. Thus, for maize (W22), which has a single ear per lateral branch, the prolificacy score is 2. LSMs with standard errors for prolificacy for each of the recombinant chromosome progeny lines and controls were determined by ANOVA with line and block effects using the software package JMP version 4.0 (SAS Institute, Cary, NC). To determine if there are pleotropic effects on other traits associated with *prol11.1*, we genotyped ~200 plants of RIL MR0091 that segregates for this QTL and measured tillering, number of ear branches, spikelet (kernel) number on the top ear of the plant, and the weight of 100 kernels. Plants for these experiments were grown at the West Madison Agricultural Research Station in Madison, WI.

**Expression Assays.** For all expression assays, total cellular RNA was isolated using Trizol (Invitrogen) from immature ear-forming axillary branches. A 1 µg aliquot of each of RNA sample was DNase treated and reverse transcribed using a polyT primer and Superscript III reverse transcriptase (Invitrogen). cDNA integrity was checked by using 0.5 µl of the RT reactions as the template for PCR (*Taq* Core Kit, Qiagen) with actin primers (ccaaggccaacagagagaaa, ccaaacggagaatagcatgag). The same actin primers were used to check for genomic DNA contamination; none was detected.

To confirm the intron-exon structure of *gt1*, PCRs were performed with cDNAs with primers (acaggctacagaggcagagc, gcgcacttgcatgataatccacac) that amplify most of the predicted transcript (Fig. S2). cDNAs derived from both the maize and teosinte alleles were used. PCR products were assayed on standard Tris-borate-EDTA agarose gels. These PCRs consistently revealed three size classes of products for both maize and teosinte alleles. These PCR products were cloned using TOPO® TA Cloning Kit (Invitrogen) and the clones sequenced at the University of Wisconsin Biotechnology Center using Sanger sequencing. Since the relative abundance of the three PCR size classes differed between the maize and teosinte alleles, we also assayed cDNAs derived from two lines with recombinant alleles: one having teosinte "causative region" and maize coding region (W22-QTL1S-IN0383), the other having maize "causative region" and teosinte coding region (W22-QTL1S-IN1043) (Fig. S1).

To compare *gt1* transcript accumulation for the maize and teosinte alleles, we performed an allele specific expression assay (10) with cDNAs from ea-forming axillary branches of 20 plants that were heterozygous for the maize/teosinte alleles of our mapping population. One µl aliquots of the 20 RT reactions were used as the template for PCRs with a primer pair in the 3' UTR of *gt1* including one fluorescently labeled primer (5'-FAM-catgatggacctcgcgcccg, gcgcacttgcatgataatccacac). This primer pair flanks a 2 bp indel that distinguishes the maize and teosinte transcripts. PCR products were assayed on an ABI 3700 fragment analyzer (Applied Biosystems) and the areas under the peaks corresponding to the maize and teosinte transcripts were determined using Gene Marker version 1.70. The relative message level associated with the maize vs. teosinte alleles in each of the twenty samples was calculated as the ratio of the area under teosinte/maize allele peaks. Two technical replicates were performed for each of the 20 biological replicates. The same assay was also performed with the DNA from each plant used for RNA extraction to assess any bias in allele amplification in the PCRs. The DNA analysis showed a slight bias towards the maize allele with maize/teosinte ratios of 1.05. Thus, the area under the teosinte peak with the cDNAs was multiplied by 1.05 to correct this bias.

We also compared transcript accumulation for the maize and teosinte alleles using quantitative real-time PCR (qPCR) with cDNA from immature ear-forming axillary branches of 12 homozygous maize and 12 homozygous teosinte plants as described above. For this assay, cDNA was first concentrated using RNAClean XP beads (Beckman Coulter). qPCR was performed on ABI Prism 7000 sequence detection system (Applied Biosystems) with Power SYBR Green PCR Master Mix (Applied Biosystems). Transcript abundance for *gt1* was assayed using a set of primers in the 3' UTR (gcaatcaaggtcactagtatagtctg; gcgcacttgcatgataatccacac). Actin primers (see above) were used as the control. The annealing temperature/time used were 52°C for 30 sec; the extension temperature/time were 72°C for 45 sec.

**In situ hybridization.** Young ear-forming axillary buds (44-50 days after planting) were collected from the top two nodes bearing lateral buds from field grown plants. These ears were fixed in 4% para-formaldehyde 1 X phosphate-buffered saline overnight at 4 $^0$C, then dehydrated with an ethanol series and embedded in paraffin wax. Embedded tissue was sectioned to 8μM with a Leica RM2155 microtome. The full *gt1* cDNA coding sequence was used as a probe as described previously (9). In situ hybridization with digoxygenin-UTP labeled antisense probe was preformed as previously described (54). Strong *gt1* expression characteristic of dormant lateral bud leaves or tassel floret carpels requires a relatively short development of the color reaction (3-4 hrs), while weaker *gt1* expression in leaves of non-dormant buds and shoot nodes requires a more extended development (15-20 hrs.).

**Population Genetics.** We sequenced the *gt1* control region plus some flanking sequence (AGP v2: 23,231,760 to 23,235,500) for a set of 15 diverse maize and 9 diverse teosinte lines (Table S4). Initial PCR primers were designed at either end of this interval based on the B73 reference genome. PCR products for each of the 24 diverse lines were sequenced using the Sanger method. A primer walk across the interval was performed for each of the 24 lines. In cases where B73 specific primers failed for one of the diverse lines because of sequence divergence or large insertions, we used consensus sequence data from the diverse lines that were successfully amplified to design primers in conserved regions.

Sequences were aligned with Clustal X (55), and checked manually. Alignment regions with gaps or ambiguous alignment were removed from further analysis. Because the teosinte and maize individuals sequenced were inbred lines, we treated the sequence as haploid data (Table S4). After removing all gapped and tri-allelic sites, 2,871 base pairs remained. We calculated the number of segregating sites ($S$), nucleotide diversity ($\pi$) and Tajima's $D$ for both maize and teosinte using custom perl scripts. We used MEGA5 (56) to infer a neighbor-joining (NJ) tree for the region (Fig. S4A), and STRUCTURE (57) to test for admixture (*SI Appendix*). We used structured coalescent simulations to estimate the maximum likelihood values of the selection coefficient (s) and degree of dominance (h) of the class-*M* haplotype. We simulated a simple domestication model including a demographic bottleneck and a partial selective sweep (*SI Appendix*). Coalescent simulations made use of a modified version of the mbs software (58).

To estimate population frequencies of the class-*M* and class-*T* haplotypes in the *gt1* control region, we chose an ~250 bp insertion in the teosinte haplotype at AGP v2: 23,232,564 in the B73 reference genome as a marker for the teosinte haplotype. This insertion was identified from the sequences of the 24 diversity lines discussed above. The insertion is present all of the class-*T* haplotypes. Primers (gagactggcgactggtcct, gacgtgcagacagcagacat) were designed in conserved sequences flanking the insertion. PCRs with these primers yield an ~600 bp product for the teosinte haplotype and an ~350 bp product for the maize haplotype. PCR product size differences were scored on 2% agarose gels for a panel of 68 maize landraces, 90 *Z. mays* ssp. *parviglumis* and 96 *Z. mays* ssp. *mexicana* (Table S5).

**Acknowledgments**

We thank Madelaine Bartlett and Graham Coop for helpful discussion. This work was supported by the USDA-Hatch grant MSN101593, USDA-NIFA grant 2009-01864, the National Science Foundation grants IOS1025869 and IOS0820619, and start-up funds to CJW from Brigham Young University.

## Literature Cited

1. Hammer K (1984) Das domestikationssyndrom. [The domestication syndrome] *Kulturpflanze* 32:11-34 (German).

2. Harlan JR (1992). Crops and man (Madison, Wisconsin: American Society of Agronomy).

3. Gepts P (2002) A Comparison between crop domestication, classical plant breeding, and genetic engineering. *Crop Sci* 42:1780-1790.

4. Matsuoka Y, et al. (2002) A single domestication for maize shown by multilocus microsatellite genotyping. *Proc Natl Acad USA* 99:8060-8064.

5. Piperno DR, Ranere AJ, Holst I, Iriarte J, and Dickau R (2009) Starch grain and phytolith evidence for early ninth millennium BP maize from the Central Balsas River Valley, Mexico. *Proc Natl Acad USA* 106:5019-5024.

6. Iltis HH (1983) From teosinte to maize: The catastrophic sexual transmutation. *Science* 222:886-94.

7. Doebley JF, Stec A, Hubbard L (1997) The evolution of apical dominance in maize. *Nature* 386:485-488.

8. Shannon LM (2012) The genetic architecture of maize domestication and range expansion. PhD Thesis, University of Wisconsin-Madison.

9. Whipple CJ, et al. (2011). grassy tillers1 promotes apical dominance in maize and responds to shade signals in the grasses. *Proc Natl Acad USA* 108:e506-e512.

10. Clark R, Nussbaum-Wagler T, Quijada P, Doebley J (2006) A distant upstream enhancer at the maize domestication gene, *tb1*, has pleiotropic effects on plant and inflorescent architecture. *Nat Genet* 38:594-597.

11. Arbor, A (1930) Studies in the Gramineae IX. 1. The nodal plexus. 2. Amphivasal bundles. *Annals of Botany* 44:593-620.

12. Wright S, et al. (2005) The effects of artificial selection on the maize genome. *Science* 308:1310-1314.

13. Eyre-Walker A., Gaut RL, Hilton H, Feldman DL, Gaut BS (1998) Investigation of the bottleneck leading to the domestication of maize. *Proc Natl Acad USA* 95:4441-4446.

14. Hudson RR (2007) The variance of coalescent time estimates from DNA sequences. *J. Mol. Evol.* 64:702-705.

15.    Thomson R, Pritchard JK, Shen PD, Oefner PJ, Feldman MW (2000) Recent common ancestry of human Y chromosomes: Evidence from DNA sequence data. *Proc Natl Acad USA* 97:7360-7365.

16.    Paterniani E, Goodman MM, (1977) Races of maize in Brazil and adjacent areas. CIMMYT, Mexico City.

17.    Doebley JF, Wendel JF, Smith JSC, Stuber CW, Goodman MM (1988) The origin of cornbelt maize: the isozyme evidence. *Econ Bot* 42:120-131.

18.    Tanksley SD (2004) The genetic, developmental and molecular bases of fruit size an shape variation in tomato.  *Plant Cell* 16:S181-S189.

19.    Doebley J, Stec A (1991) Genetic analysis of the morphological differences between maize and teosinte. *Genetics* 129:285-295.

20.    Doebley JF, Stec A (1993) Inheritance of the morphological differences between maize and teosinte: comparison of results for two $F_2$ populations. *Genetics* 134:559-570.

21.    Briggs WH, McMullen MD, Gaut BS, Doebley J (2007)   Linkage mapping of domestication loci in a large maize teosinte backcross resource. *Genetics* 177:1915-1928.

22.    Grandillo S, Tanksley SD (1996) QTL analysis of horticultural traits differentiating the cultivated tomato from the closely related species *Lycopersicon pimpinellifolium*.  *Theor Appl  Genet*  92:935-952

23.    Doganlar S, Frary A, Ku HM, Tanksley SD (2002) Mapping quantitative trait loci in inbred backcross lines of *Lycopersicon pimpenellifolium* (LA1589). *Genome* 45:1189-1202.

24.    Koinange EMK, Singh SP, Gepts P (1996)  Genetic control of the domestication syndrome in common bean.  *Crop Science* 36:1037-1045.

25.    Poncet V, et al. (2000) Genetic control of domestication traits in pearl millet (*Pennisetum glaucum* L., Poaceae). *Theor Appl Genet* 100:147-159.

26.    Wills DM, Burke JM (2007)  Quantitative trait locus analysis of the early domestication of sunflower.  *Genetics*  176:2589-2599.

27.    Tang S, Leon A, Bridges WC, Knapp SJ *(*2006)  Quantitative trait loci for genetically correlated seed traits are tightly linked to branching and pericarp pigment loci in sunflower. *Crop Science* 46:721-734.

28.    Lester RN (1989)  Evolution under domestication involving disturbance of genic balance . *Euphytica* 44:125-132.

29. Studer A, Zhao Q, Ross-Ibarra J, Doebley J (2011) A transposon insertion was the causative mutation in the maize domestication gene *tb1*. *Nat Genet* 43:1160-1163.

30. Wang H, et al. (2005). The origin of the naked grains of maize. *Nature* 436:714-719.

31. Beadle GW (1939) Teosinte and the origin of maize. *J Hered* 30:245-247.

32. Doebley JF (2004) The genetics of maize evolution. *Annual Review of Genetics* 38:37-59.

33. Studer A, Doebley J (2011) Do large effect QTL fractionate? A case study at the maize domestication QTL *teosinte branched1*. *Genetics* 188:673-681

34. Gross BL, Olsen KM (2010) Genetic perspectives on crop domestication. *Trends in Plant Science* 15:529-537.

35. Lin Z., et al. (2012) Parallel domestication of the *Shattering1* genes in cereals. *Nat Genet* 44:720-724.

36. Li C, Zhou A, Sang T (2006) Rice domestication by reducing shattering. *Science* 311:1936-1939.

37. Konishi S, et al. (2006) An SNP caused loss of seed shattering during rice domestication. *Science* 312:1392-1396.

38. Cong B, Barrero LS, Tanksley SD (2008) Regulatory change in YABBY-like transcription factor led to evolution of extreme fruit size during tomato domestication *Nat Genet* 40:800-804.

39. Frary A, et al. (2000) *fw2.2*: A quantitative trait locus key to the evolution of tomato fruit size. *Science* 289:85-88.

40. Jin J, et al. (2008) Genetic control of rice plant architecture under domestication. *Nat Genet* 40:1365-1369.

41. Tan L, et al. (2008) Control of a key transition from prostrate to erect growth in rice domestication. *Nat Genet* 40:1360-1364.

42. Hung HY, et al. (2012) *ZmCCT* and the genetic basis of day-length adaptation underlying the post-domestication spread of maize. *Proc Nat Acad USA* 109:E1913-E1921.

43. Komatsuda, T, et al. (2007) Six-rowed barley originated from a mutation in a homeodomain-leucine zipper I-class homeobox gene. *Proc Natl Acad USA* 104:1424-1429.

44.  Dussert Y, et al. (2013) Polymorphism pattern at a miniature inverted-repeat transposable element locus downstream of the domestication gene *Teosinte-branched1* in wild and domesticated pearl millet. *Mol Ecol* 22: 327-340.

45.  Xiao H., et al. (2008) A Retrotransposon-Mediated Gene Duplication Underlies Morphological Variation of Tomato Fruit. *Science* 319 1527-1530.

46.  Gallavotti A, et al. (2004) The role of *barren stalk1* in the architecture of maize. *Nature* 432:630-635.

47.  Ralph P, Coop G  (2010) Parallel adaptation: one or many waves of advance of an advantageous allele? *Genetics* 186:647-668.

48.  Hufford MB, et al.  (2012)  Comparative population genomics of maize domestication and improvement.  *Nat Genet* 44:808-811.

49.  Innan H, Kim Y (2004) Pattern of polymorphism after strong artificial selection in a domestication event. *Proc Natl Acad USA* 101:10667-10672.

50.  Hermisson J, Pennings PS (2005) Soft sweeps: molecular population genetics of adaptation from standing genetic variation. *Genetics* 169:2335-2352.

51.  Smith H,  Whitelam GC (1997) The shade avoidance syndrome: multiple responses mediated by multiple phytochromes.  *Plant, Cell and Environment* 20:840-844.

52.  Elshire RJ, et al.  (2011) A robust, simple genotyping-by-sequencing (GBS) approach for high diversity species. *PLoS ONE* 6:e19379.

53.  Broman KW, Wu H, Sen S, Churchill GA (2003) R/qtl: QTL mapping in experimental crosses. *Bioinformatics* 19:889-890.

54.  Jackson D, Veit B, Hake S (1994)  Expression of maize KNOTTED1 related homeobox genes in the shoot apical meristem predicts patterns of morphogenesis in the vegetative shoot.  *Development* 120:405-413.

55.  Larkin MA, et al.  (2007). Clustal W and Clustal X version 2.0. *Bioinformatics*, 23:2947-2948.

56.  Tamura K, et al. (2011) MEGA5: Molecular evolutionary genetics analysis using maximum likelihood, evolutionary distance, and maximum parsimony methods. *Mol Biol Evol* 28:2731-2739.

57.  Pritchard JK, Stephens M, Donnelly P, (2000) Inference of  population structure using multilocus genotype data. *Genetics* 155:945-959.

58. Teshima KM, Innan H (2009)  mbs: modifying Hudson's ms software to generate samples of DNA sequences with a biallelic site under selection. *BMC Bioinformatics* 10:166.

**Table 1.  Summary of QTL for prolificacy in a set of Maize-Teosinte BC₂S₃ RILs**.  The physical and genetic positions of the 1.5 LOD support intervals for each QTL are shown in addition to the physical and genetic size of the intervals. LOD scores and percent variance explained (PVE) as calculated by the fitqtl function in R/qtl, which performs a drop-one ANOVA, are reported. Physical positions are reported in Mbp along the chromosome for the maize reference genome (Maize Reference Genome AGP v2).

| Chr | QTL | Left physical position | Right physical position | 1.5 LOD Support Interval (cM) | 1.5 LOD Support Interval (Mb) | LOD | PVE |
|---|---|---|---|---|---|---|---|
| 1 | *prol1.1* | 22.63 | 23.42 | 1.62 | 0.79 | 157.23 | 36.7 |
| 1 | *prol1.2* | 179.93 | 180.31 | 0.59 | 0.38 | 5.68 | 0.86 |
| 1 | *prol1.3* | 182.17 | 202.55 | 2.59 | 20.38 | 8.32 | 1.27 |
| 2 | *prol2.1* | 44.26 | 62.99 | 3.33 | 18.73 | 7.32 | 1.11 |
| 3 | *prol3.1* | 196.09 | 197.40 | 1.17 | 1.31 | 18.01 | 2.82 |
| 4 | *prol4.1* | 157.62 | 165.70 | 6.48 | 8.08 | 15.05 | 2.34 |
| 4 | *prol4.2* | 195.96 | 200.74 | 4.52 | 4.78 | 7.89 | 1.20 |
| 5 | *prol5.1* | 140.09 | 145.34 | 1.24 | 5.25 | 36.71 | 6.05 |

**Table 2.** Frequencies of a ~250 bp insertion in the teosinte haplotype of the *gt1* causative region in a diverse sample of maize and teosinte.

|  | Sample Size | Insertion - | Insertion + |
|---|---|---|---|
| Maize landraces | 68 | 0.706 | 0.294 |
| ssp. *parviglumis* | 90 | 0.050 | 0.950 |
| ssp. *mexicana* | 96 | 0.078 | 0.922 |

**Figure Legends**

**Fig. 1.** Prolificacy phenotypes. (A) Segment of a teosinte lateral branch showing a cluster of ears at the node. Three of the ears still have their husk leaf around them. (B) Side branch of one of our isogenic lines carrying the teosinte allele at *prol1.1* and showing the cluster of ears that this allele engenders. (C) Side branch of one of our isogenic lines carrying the maize allele at *prol1.1* and showing a single terminal ear as is typical for maize.

**Fig. 2.** LOD plots from a genome wide QTL scan for prolificacy in a set of maize-teosinte $BC_2S_3$ RILs. Densely spaced black hash marks along the bottom axis represent genetic markers, curves represent logarithm of odds (LOD) scores for QTL at each genomic position. LOD curves for distinct QTLs on a single chromosome are plotted with different colors. The dotted horizontal line at LOD 4.44 represents the threshold for significance as determined by 10,000 permutations of the data.

**Fig. 3.** Fine-mapping of *prol1.1* on chromosome 1S. At the top, there is a map of the *prol1.1* chromosomal region with genetic markers and their APG v2 positions. The upper set of 25 horizontal bars represents the 23 recombinant chromosome lines and the maize and teosinte control lines. White segments indicate maize genotype, black segments teosinte genotype, and gray segments unknown or regions where maize and teosinte are identical. Prolificacy trait values and standard errors for each recombinant and control line are shown by the blue column graphs on the right. The lower set of 25 bars is a close-up view of the region near *gt1* to which *prol1.1* localized. At the bottom, a fine-scale map showing the location of *prol1.1* between SBM07 and SBM08 and its position relative to the *gt1* coding sequence. See also Fig. S1 and Table S1.

**Fig. 4.** Agarose gel image showing RT-PCR products for *gt1*. Lanes show the maize (W22) allele, teosinte allele, recombinant allele with the maize control region and teosinte coding region (M:T), and recombinant allele with the teosinte control region and maize coding region (T:M). The outer two lanes are molecular size markers with the sizes in bp indicated.

**Fig. 5.** Longitudinal sections of ear-forming primary lateral branches hybridized with antisense *gt1* RNA probe. (A) M:M and (B) M:T genotypes, showing *gt1* expressed at low levels in the nodes. (C) T:M and (D) T:T genotypes in which there is no viable *gt1* expression in the nodes. Weak *gt1* expression is seen in the leaves surround the branch in all sections.

**Fig. 6.** (A) Pattern of pairwise linkage disequilibrium ($r^2$) between 67 SNPs in maize landraces, including 13 class-*M* and 2 class-*T* haplotypes. (B) Likely demographic model of the process of maize domestication (see text and *SI Appendix* for details). The ancestral (wild) and current population sizes of maize are denoted by $N_A$ and $N_P$, respectively. The domestication bottleneck started at $t_d$ generations ago, and ended at $t_e$ generations ago. $N_B$ and $t_B$ represent the size and duration of the bottleneck, respectively. The trajectory of class-*M* haplotype is shown by dashed (neutral in the wild population) and solid (positively selected after domestication) lines. *f* represents the current frequency of class-*M* haplotype in maize. (C) Heat map of the maximum likelihood estimates of the intensity of selection (*s*) and the degree of dominance (*h*). The likelihood given *s* and *h* is denoted by $L(s, h)$, and the scale bar is also shown.

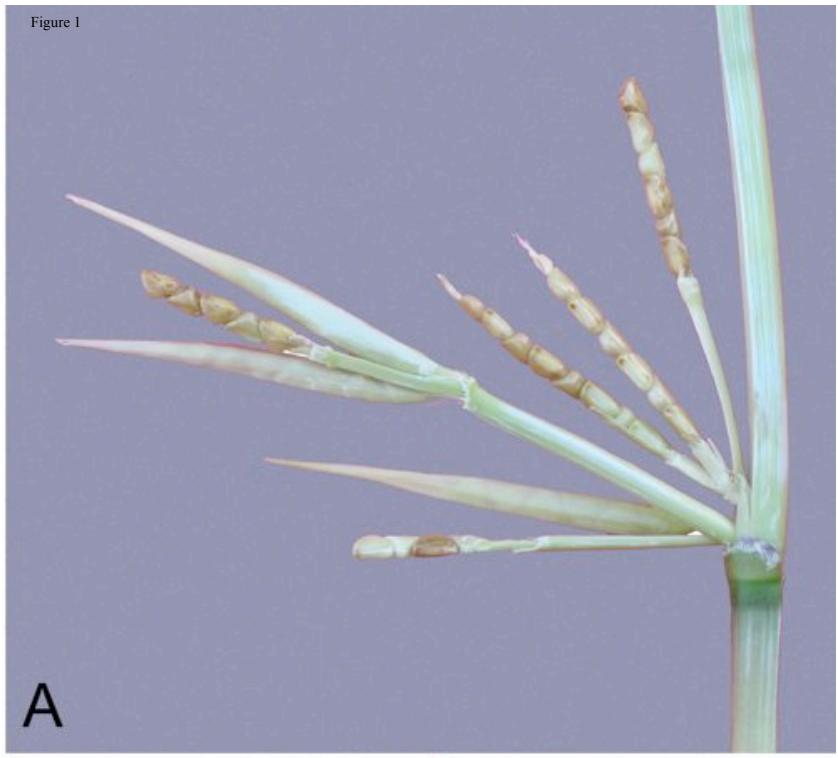

A

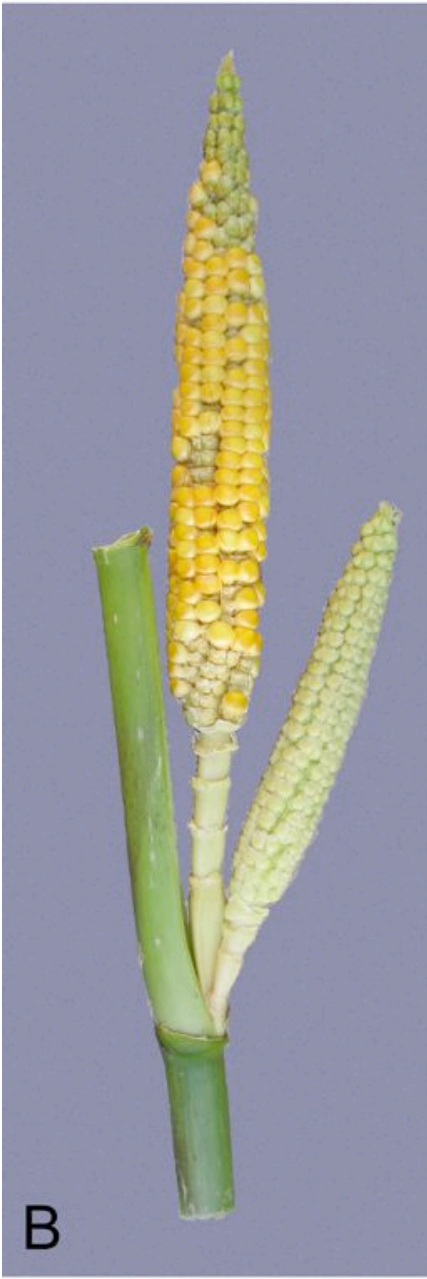

B

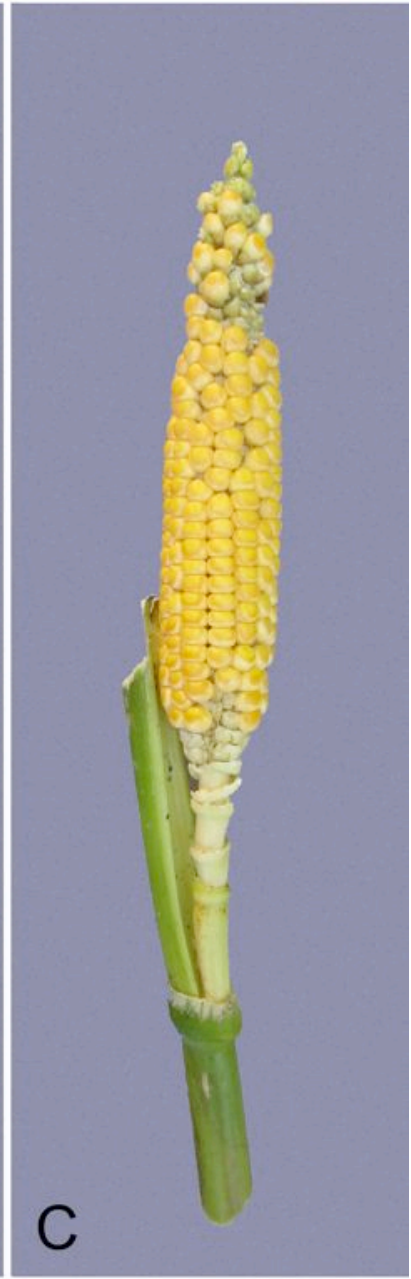

C

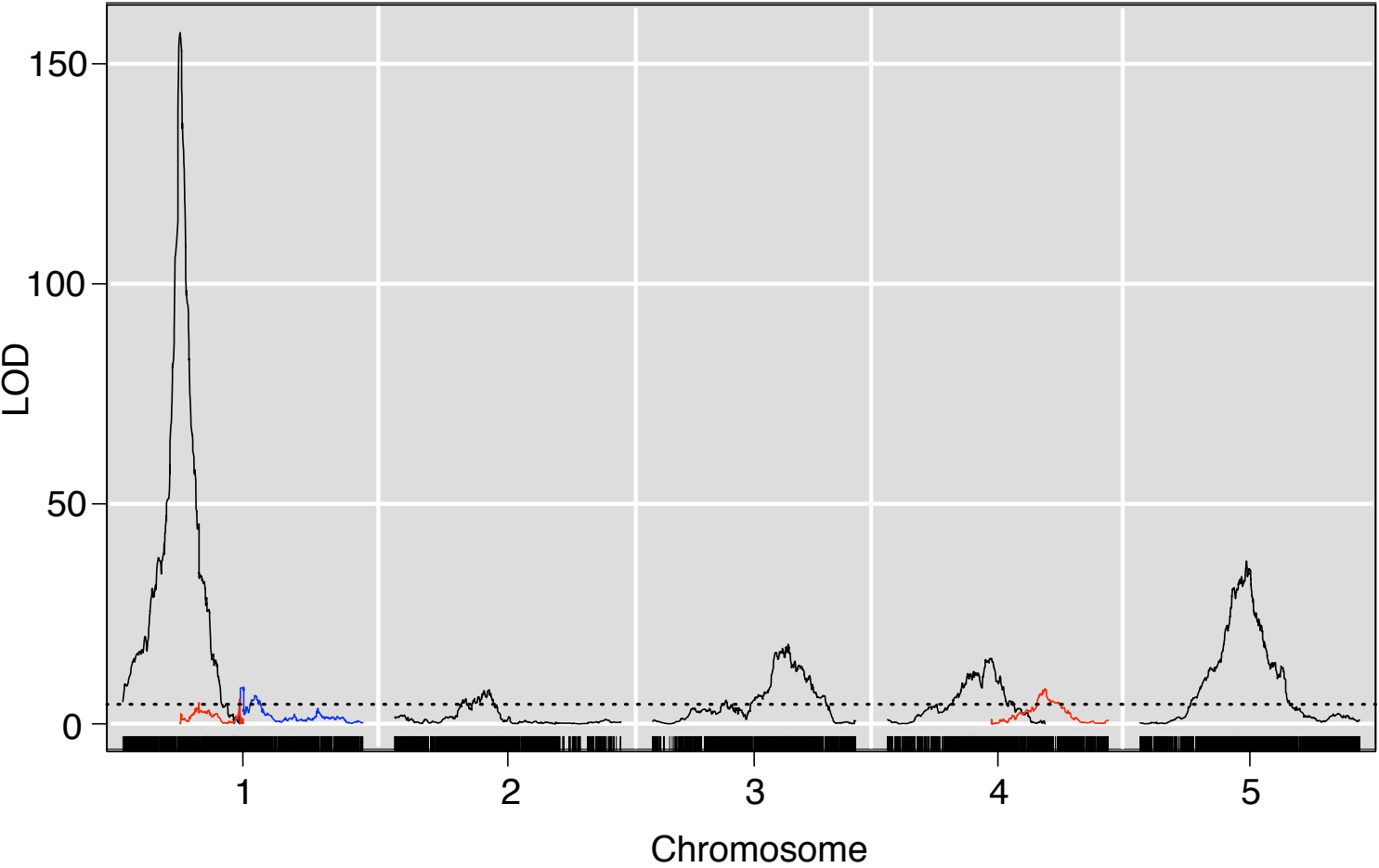

Figure 2

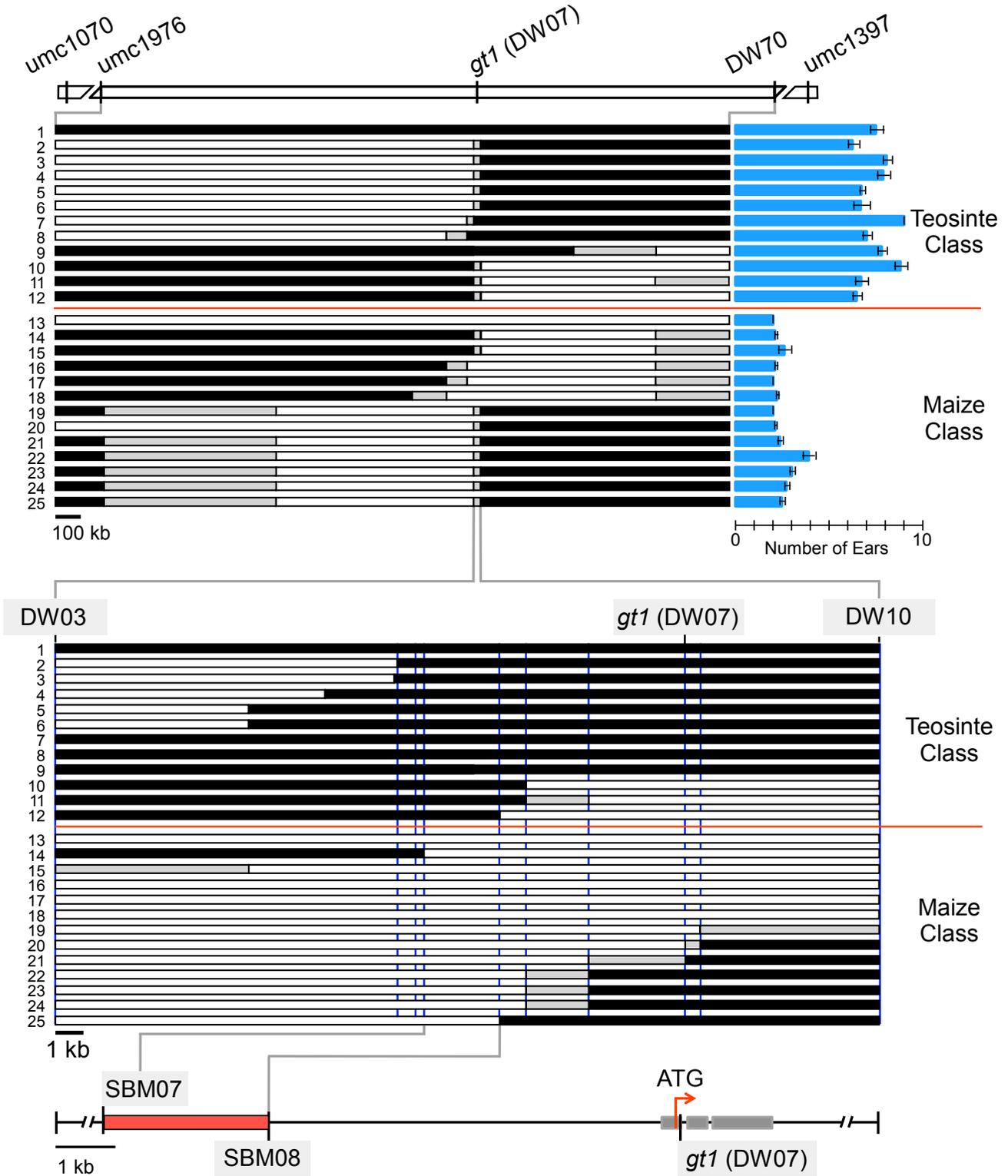

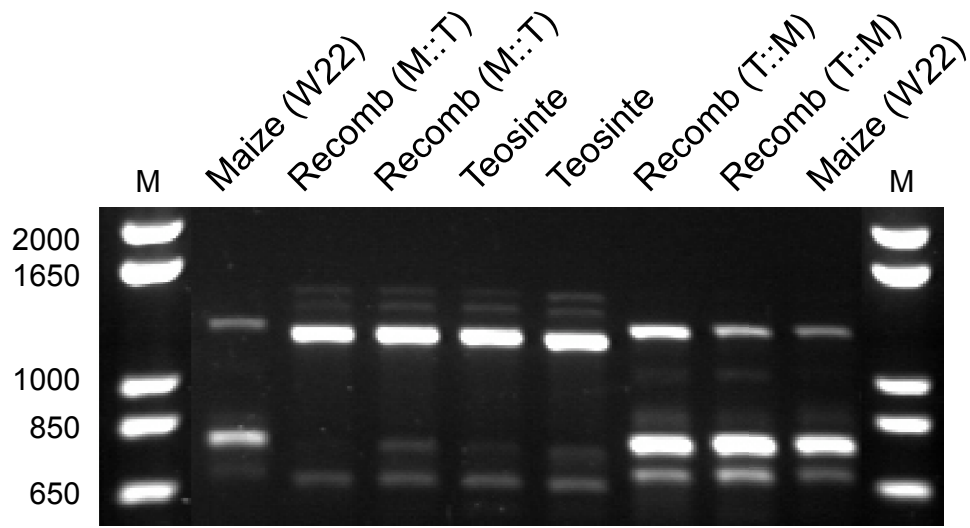

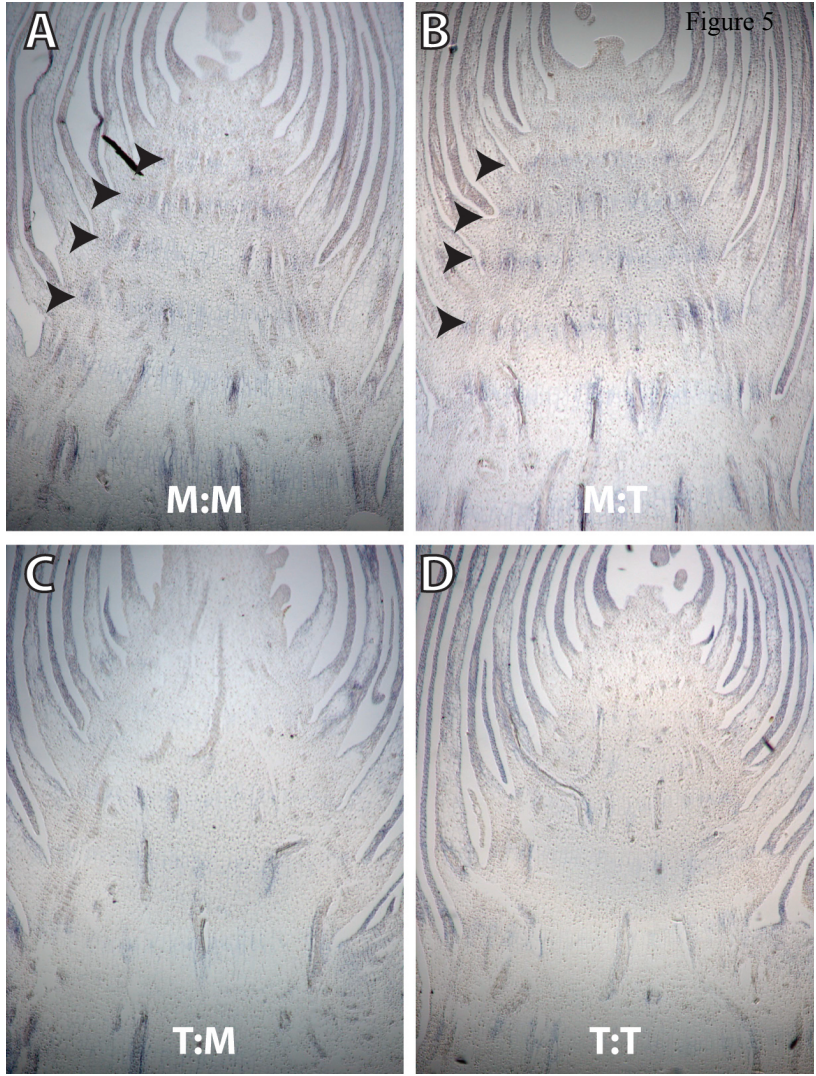

Figure 5

**A**

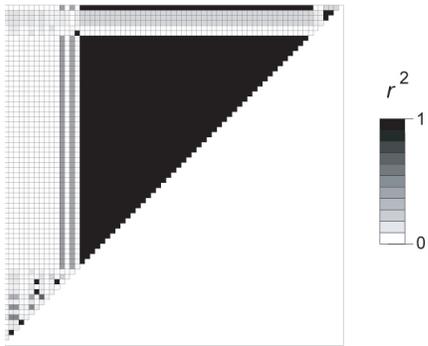

$r^2$

0

**B**

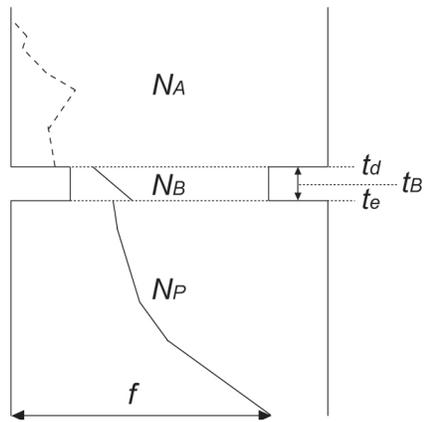

$N_A$

$N_B$

$N_P$

$t_d$

$t_B$

$t_e$

$f$

**C**

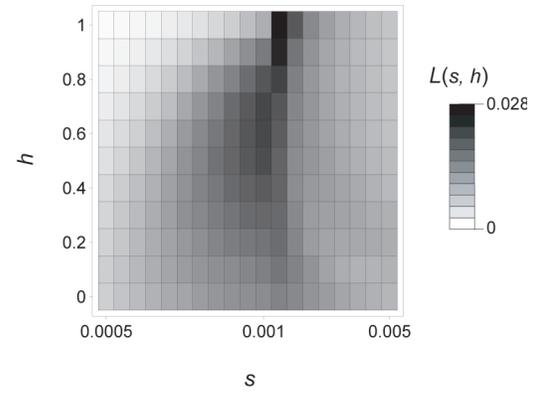

$L(s, h)$

0.028

0